\documentclass[useAMS,usenatbib]{mn2e}

\usepackage[dvips]{graphicx}
\usepackage{natbib}

\newcommand{\teff}{\ensuremath{T_{\rm eff}}}

\newcommand{\kms}{km\,s$^{-1}\,$}

\newcommand{\vinf}{$v_\infty$}
\newcommand{\bz}{\ensuremath{\langle B_z\rangle}}

\def\gtrsim{\mathrel{\hbox{\rlap{\hbox{\lower4pt\hbox{$\sim$}}}\hbox{$>$}}}}
\def\ltsim{\mathrel{\hbox{\rlap{\hbox{\lower4pt\hbox{$\sim$}}}\hbox{$<$}}}}

\title[Magnetic field and slow rotation of HD 108]{Confirming the oblique rotator model for the extremely slowly rotating O8f?p star HD 108 \thanks{Based on observations obtained at the Canada-France-Hawaii Telescope (CFHT) which is operated by the National Research Council of Canada, the Institut National des Sciences de l'Univers (INSU) of the Centre National de la Recherche Scientifique of France, and the University of Hawaii, under observing program CFHT15BC014.}}
\author[M. Shultz et al.]{M. Shultz\thanks{E-mail: }$^{1,2,3}$ and G.A. Wade$^2$
\\
$^{1}$Department of Physics, Engineering Physics and Astronomy, Queen's University, 99 University Avenue, Kingston, ON K7L 3N6, Canada\\
$^{2}$Dept. of Physics, Royal Military College of Canada, PO Box 17000 Station Forces, Kingston, ON, Canada K7K 0C6 \\
$^{3}$Department of Physics and Astronomy, Uppsala University, Box 516, Uppsala 75120 \\
}

\begin{document}

\date{Accepted . Received , in original form }

\pagerange{\pageref{firstpage}--\pageref{lastpage}} \pubyear{2015}

\maketitle

\label{firstpage}

\begin{abstract}

The O8f?p star HD 108 is implied to have experienced the most extreme rotational braking of any magnetic, massive star, with a rotational period $P_{\rm rot}$ of at least 55 years, but the upper limit on its spindown timescale is over twice the age estimated from the Hertzsprung-Russell diagram. HD 108's observed X-ray luminosity is also much higher than predicted by the XADM model, a unique discrepancy amongst magnetic O-type stars. Previously reported magnetic data cover only a small fraction ($\sim$3.5\%) of $P_{\rm rot}$, and were furthermore acquired when the star was in a photometric and spectroscopic `low state' at which the longitudinal magnetic field \bz~was likely at a minimum. We have obtained a new ESPaDOnS magnetic measurement of HD 108, 6 years after the last reported measurement. The star is returning to a spectroscopic high state, although its emission lines are still below their maximum observed strength, consistent with the proposed 55-year period. We measured \bz$=-325 \pm 45$ G, twice the strength of the 2007-2009 observations, raising the lower limit of the dipole surface magnetic field strength to $B_{\rm d} \ge 1$ kG. The simultaneous increase in \bz~and emission strength is consistent with the oblique rotator model. Extrapolation of the \bz~maximum via comparison of HD 108's spectroscopic and magnetic data with the similar Of?p star HD 191612 suggests that $B_{\rm d} > 2$~kG, yielding $t_{\rm S, max}<3$~Myr, compatible with the stellar age. These results also yield a better agreement between the observed X-ray luminosity and that predicted by the XADM model. 
\end{abstract}

\begin{keywords}
Stars : rotation -- Stars: massive -- Stars : individual : HD 108 -- Stars: magnetic fields -- Stars: winds, outflows.
\end{keywords}


\section{Introduction}

\begin{figure*}
\begin{centering}
\begin{tabular}{cc}
\includegraphics[width=8.5cm]{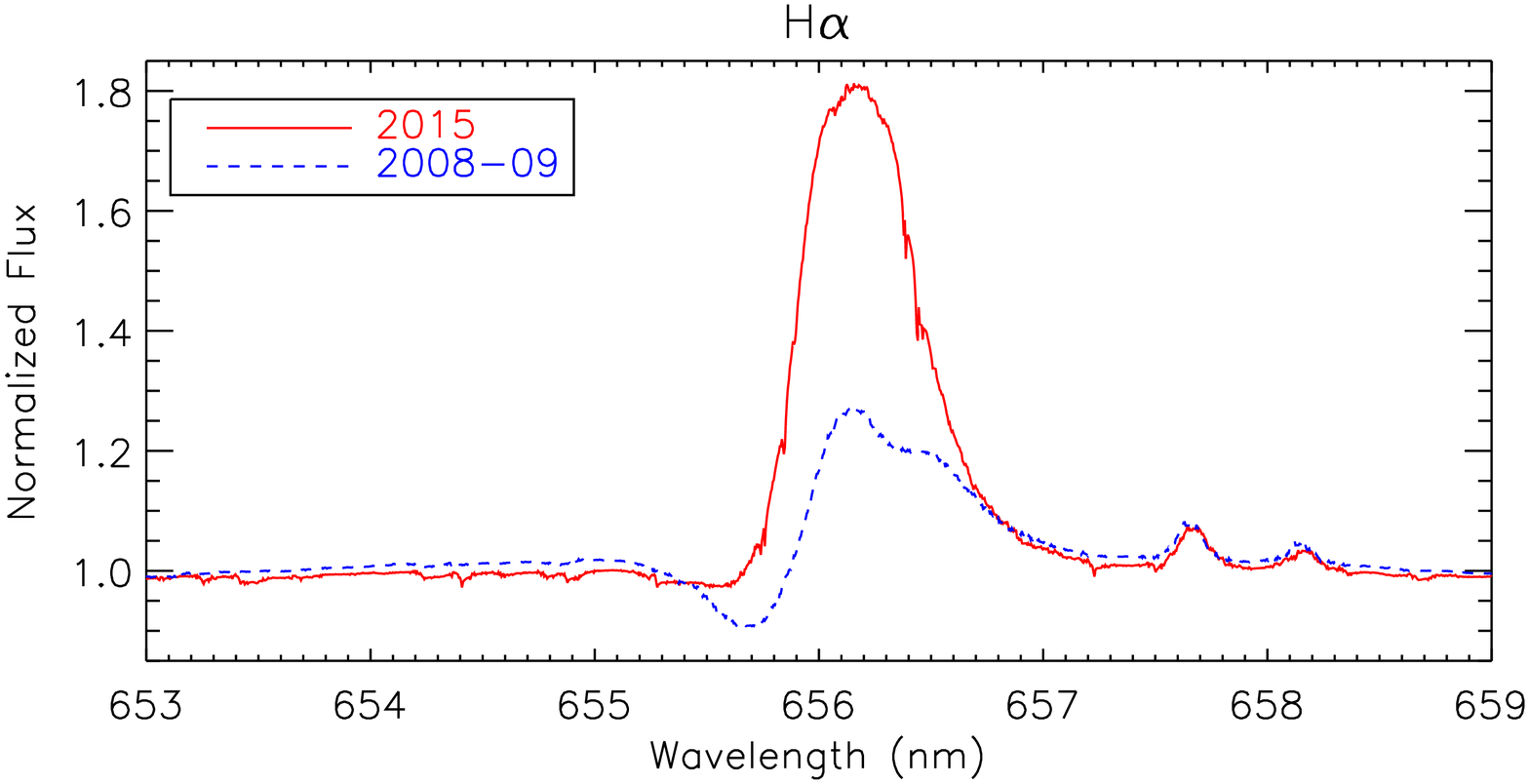} &
\includegraphics[width=8.5cm]{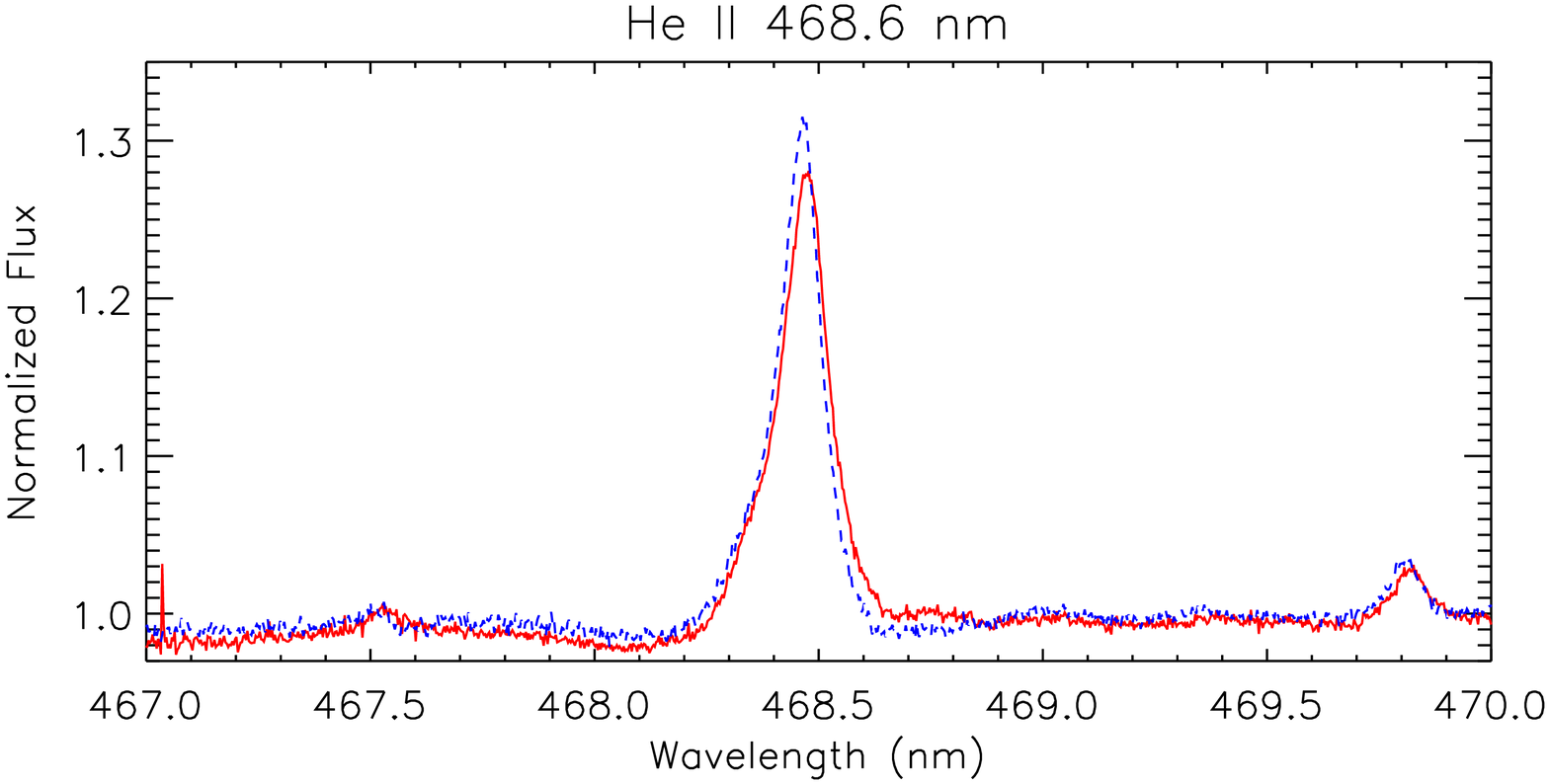} \\

\includegraphics[width=8.5cm]{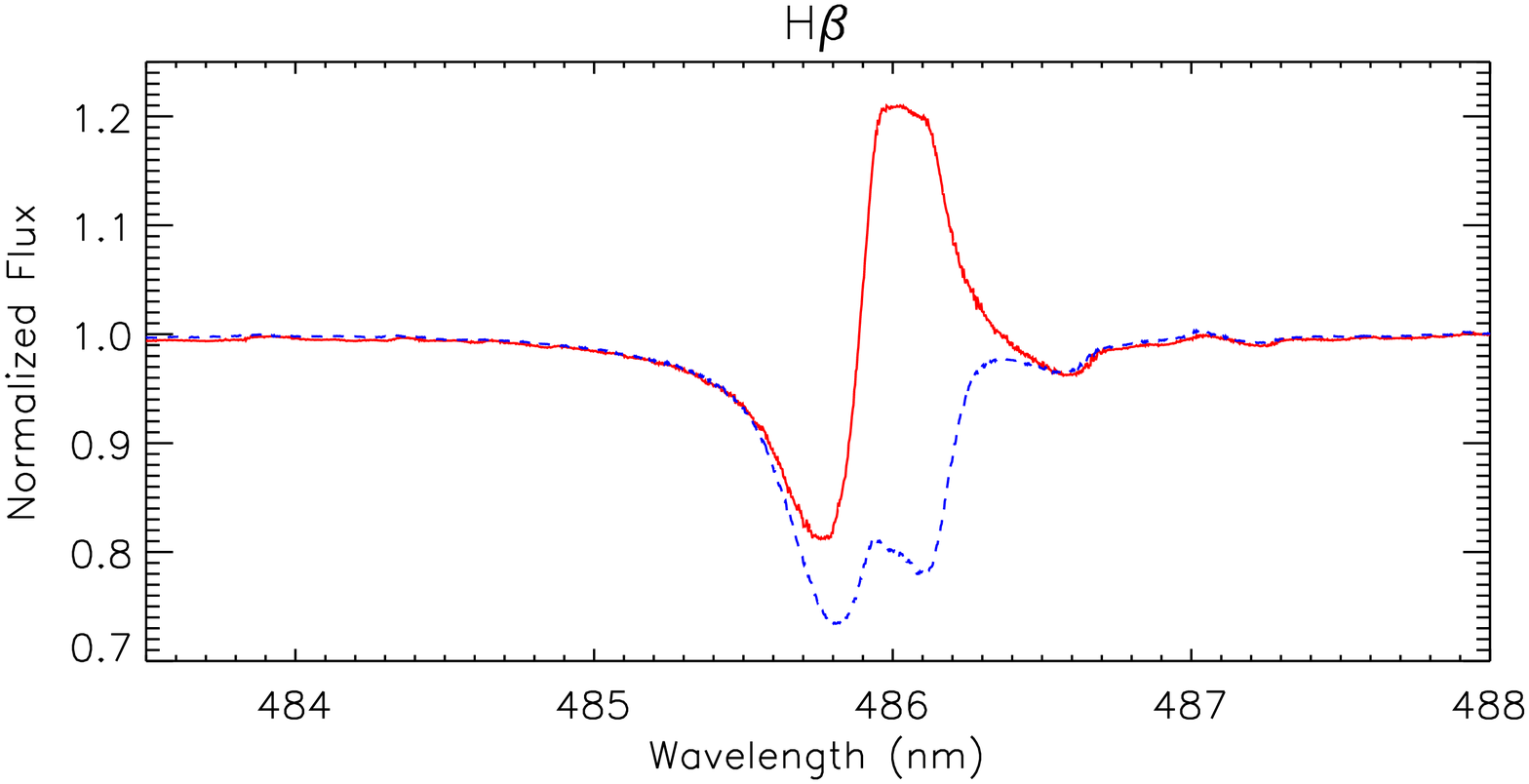} &
\includegraphics[width=8.5cm]{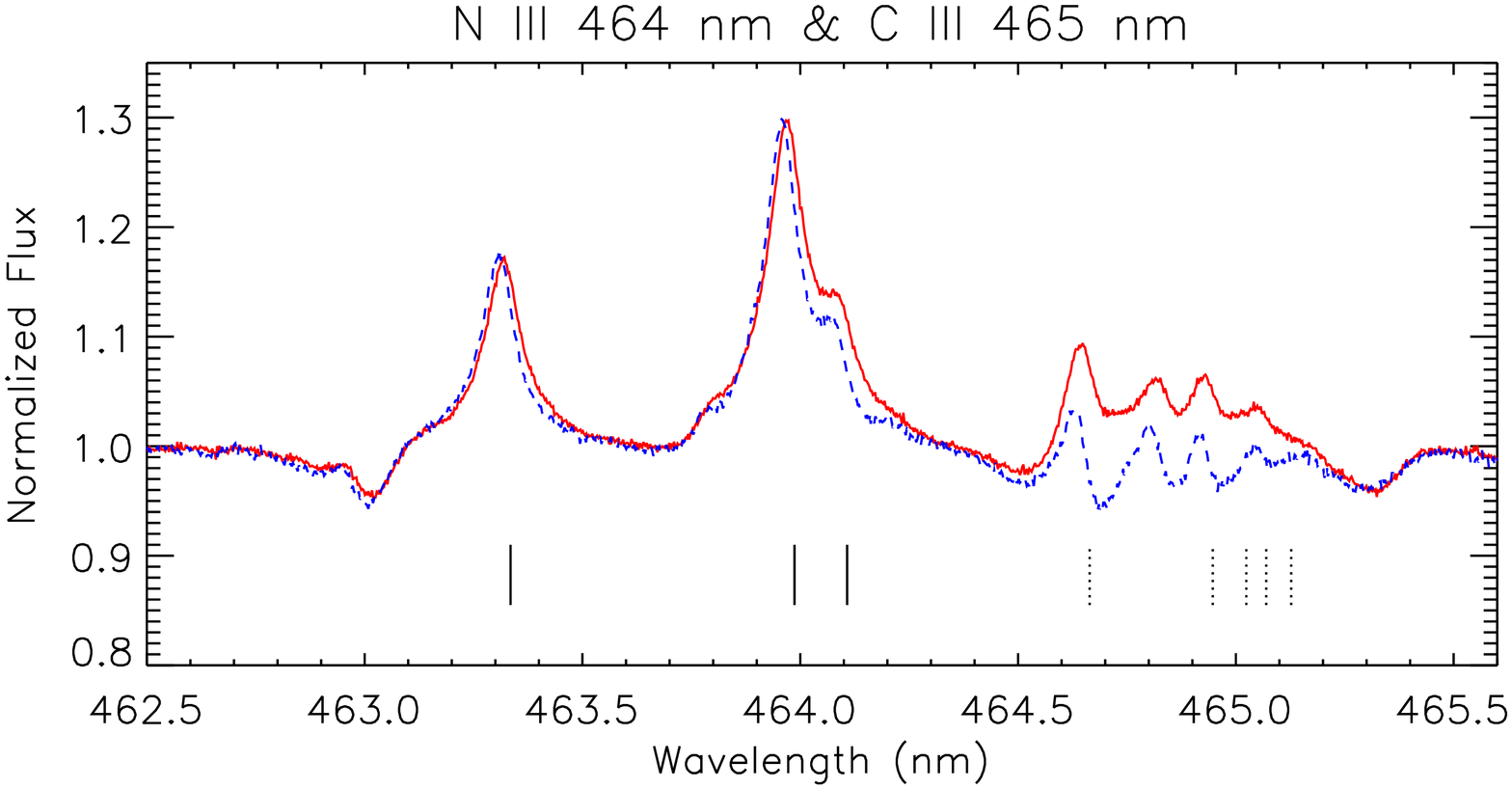} \\

\includegraphics[width=8.5cm]{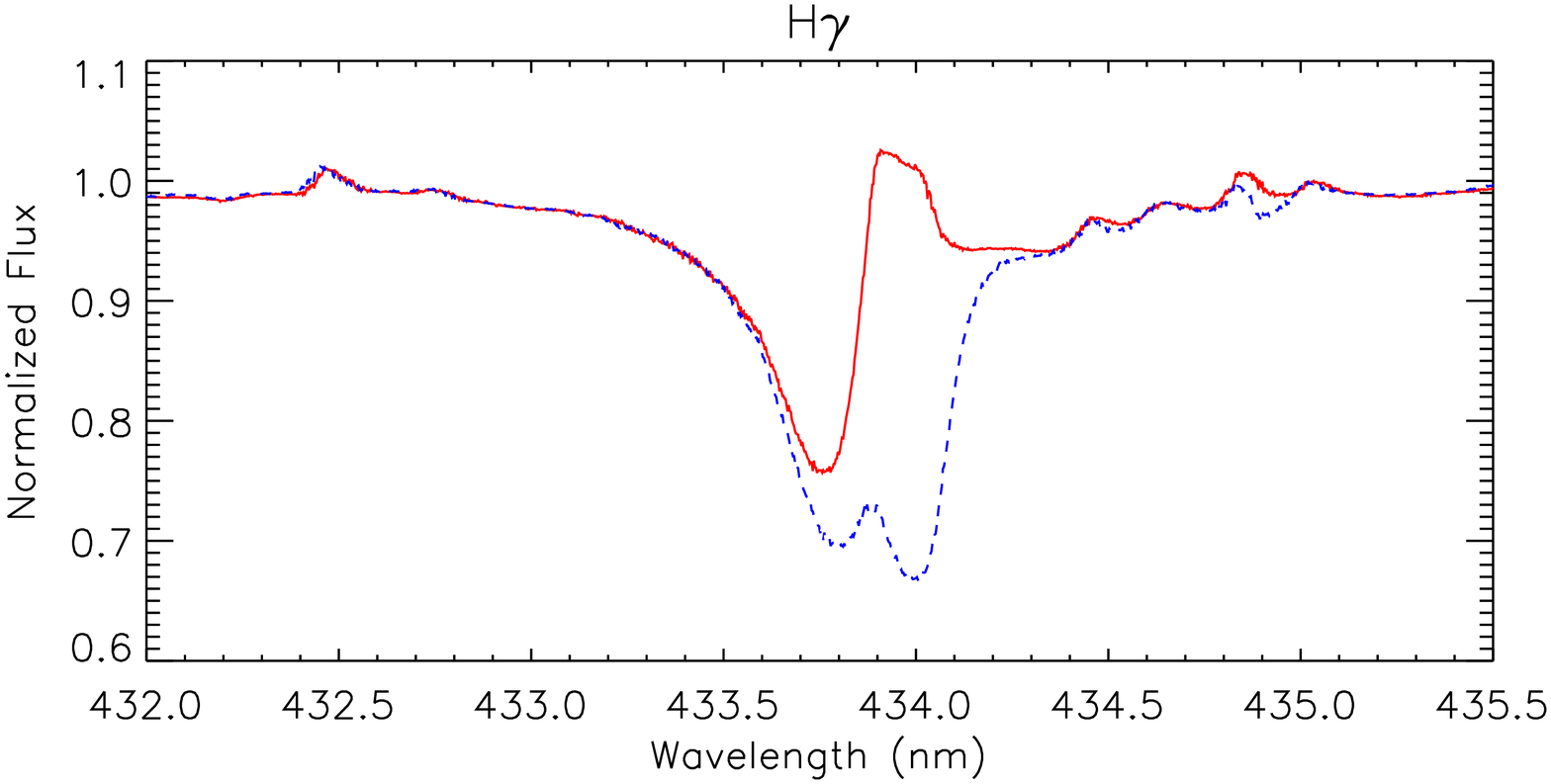} &
\includegraphics[width=8.5cm]{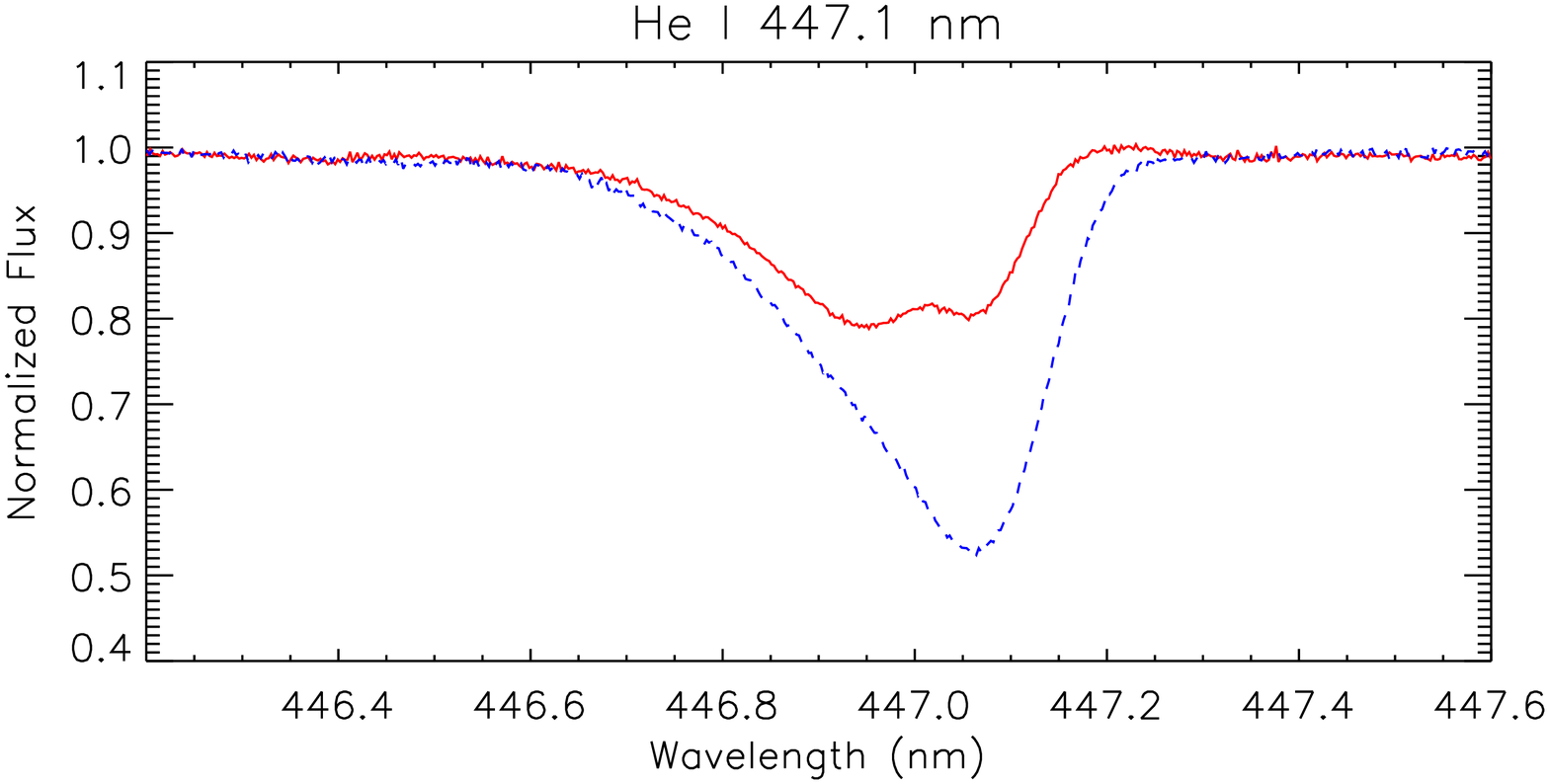} \\

\end{tabular}
\caption{{\em Left}: Comparison of H Balmer lines in 2007-2009 (dashed, blue) and 2015 (solid, red). {\em Right}: the same comparison for (top--bottom): He\,{\sc i} 447. 1 nm; the N\,{\sc iii} and C\,{\sc iii} emission lines (indicated by solid and dotted vertical lines respectively) near 464 nm; and He\,{\sc ii} 468.6 nm. In H lines and C\,{\sc iii} lines emission is much stronger in 2015 than it was in 2007-2009. He\,{\sc i} is also significantly filled in by emission; in contrast, N\,{\sc iii} and He\,{\sc ii} differ only slightly between the two epochs.}
\label{var}
\end{centering}
\end{figure*} 

The rapid spindown of magnetic stars due to angular momentum loss through their magnetized winds has been well-explored theoretically (e.g.\ \citealt{wd1967}, \citealt{ud2009}). In the case of some rapidly-rotating Bp stars with magnetic, photometric, and spectroscopic observations spanning a long temporal baseline, magnetic braking has been measured directly (e.g.\ CU Vir, \citealt{1998A&A...339..822P,2008MNRAS.384.1437T,miku2011}; $\sigma$ Ori E, \citealt{town2010}; HD 37776, \citealt{miku2008}), and the inferred timescales are in reasonably good agreement with the predictions of MHD simulations \citep{ud2009}. Magnetic braking is expected to increase with both the size of the magnetosphere, and the mass-loss rate. Consistent with this picture, magnetic OB stars are in general more slowly rotating than their non-magnetic kin, with median rotational periods of about 9 days \citep{petit2013}. Longer rotational periods of weeks and even months are frequently measured. However, there are also cases of extreme rotational braking, with apparent rotational periods on the order of decades. One star in particular is an exemplar of this class: the O8f?p star HD 108, for which evidence from spectroscopy \citep{2001AA...372..195N,2004AA...417..667N} and photometry \citep{2007IBVS.5756....1B} indicates a rotational period of between 50 and 60 years \citep{2010AA...520A..59N}. 

Magnetic measurements of HD 108 were reported by \cite{2010MNRAS.407.1423M} (hereafter M2010) for the period 2007 to 2009. They found an essentially constant longitudinal magnetic field \bz~$\sim -150$ G, from which a lower limit to the dipole surface magnetic field strength of $B_{\rm d} > 0.5$ kG was inferred. The lower limit on the spindown timescale, as inferred from a 55 yr rotation period and the lower limit on $B_{\rm d}$, is $t_{\rm S,max} = 8.5$~Myr \citep{petit2013}: this is substantially longer than the age of the star estimated from its position on the Hertzsprung-Russell diagram (HRD), $4\pm1$~Myr (M2010). An additional discrepancy is that the star's X-ray luminosity, thought to originate in magnetically confined wind shocks \citep{bm1997,ud2014}, is almost 1 dex higher than predicted, a unique occurence amongst magnetic O-type stars for which the opposite is typically the case \citep{2014ApJS..215...10N}. However, the magnetic data cover only a very small fraction ($\sim$3.5\%) of the star's presumed rotational cycle. Furthermore, the magnetic data were obtained when the star was at photometric and spectroscopic minimum. In the context of the oblique rotator model in its simplest, dipolar form, this is interpreted as a consequence of the magnetosphere being seen closest to edge-on \citep{sund2012, ud2013, 2015MNRAS.447.2551W}, corresponding to the rotational phase at which the magnetic equator bisects the stellar disk, and thus \bz~is closest to zero. It is therefore likely that $B_{\rm d}$ is substantially higher than the lower limit determined by M2010. 

In this paper we report a new ESPaDOnS observation of HD 108 that enables new constraints on the stellar rotational period and spindown timescale. In \S~2, we describe the observation. In \S~3 we examine HD 108's long-term spectroscopic variability. The magnetic analysis is presented in \S~4, and updated magnetic and magnetospheric parameters, including the spindown timescale and predicted X-ray luminosity, are determined in \S~5. In \S~6 we predict the longitudinal magnetic field variation over the full stellar rotation period, and discuss the implications of this for the star's magnetic and magnetospheric properties. Conclusions are summarized in \S~7. 

\section{Observations}

We obtained two circularly polarized (Stokes $V$) spectropolarimetric sequences of HD 108 on 2015 September 3 with ESPaDOnS, the high-dispersion ($R\sim65,000$) spectropolarimeter mounted at the 3.6 m Canada-France-Hawaii Telescope (CFHT). A detailed description of this instrument is provided by \cite{2016MNRAS.456....2W}. We followed the same strategy as that adopted by M2010: the measurement consisted of two consecutive spectropolarimetric sequences, each consisting of 4 polarized 1290~s sub-exposures, with a total exposure time of 2.9~h. Each observation yielded four unpolarized intensity (Stokes $I$) spectra, one Stokes $V$ spectrum, and two diagnostic null $N$ spectra. The data were reduced with the Upena pipeline, which incorporates the automated reduction package Libre-ESpRIT \citep{d1997}. The peak SNR per spectral pixel was 802 in the first observation, 933 in the second, and 1325 in the co-added spectrum. We have also downloaded the ESPaDOnS and Narval observations reported by M2010. Narval is a clone of ESPaDOnS mounted at the Bernard Lyot Telescope, and obtains essentially identical results to those of ESPaDOnS \citep{2016MNRAS.456....2W}. The SNR of the observations reported in this work are comparable to the mean SNR of 1295 in the nightly ESPaDOnS observations presented by M2010. The spectra were normalized by fitting polynomial splines to the continuum flux in individual orders. 

\section{Variability}

\begin{figure}
\begin{centering}
\includegraphics[width=\hsize]{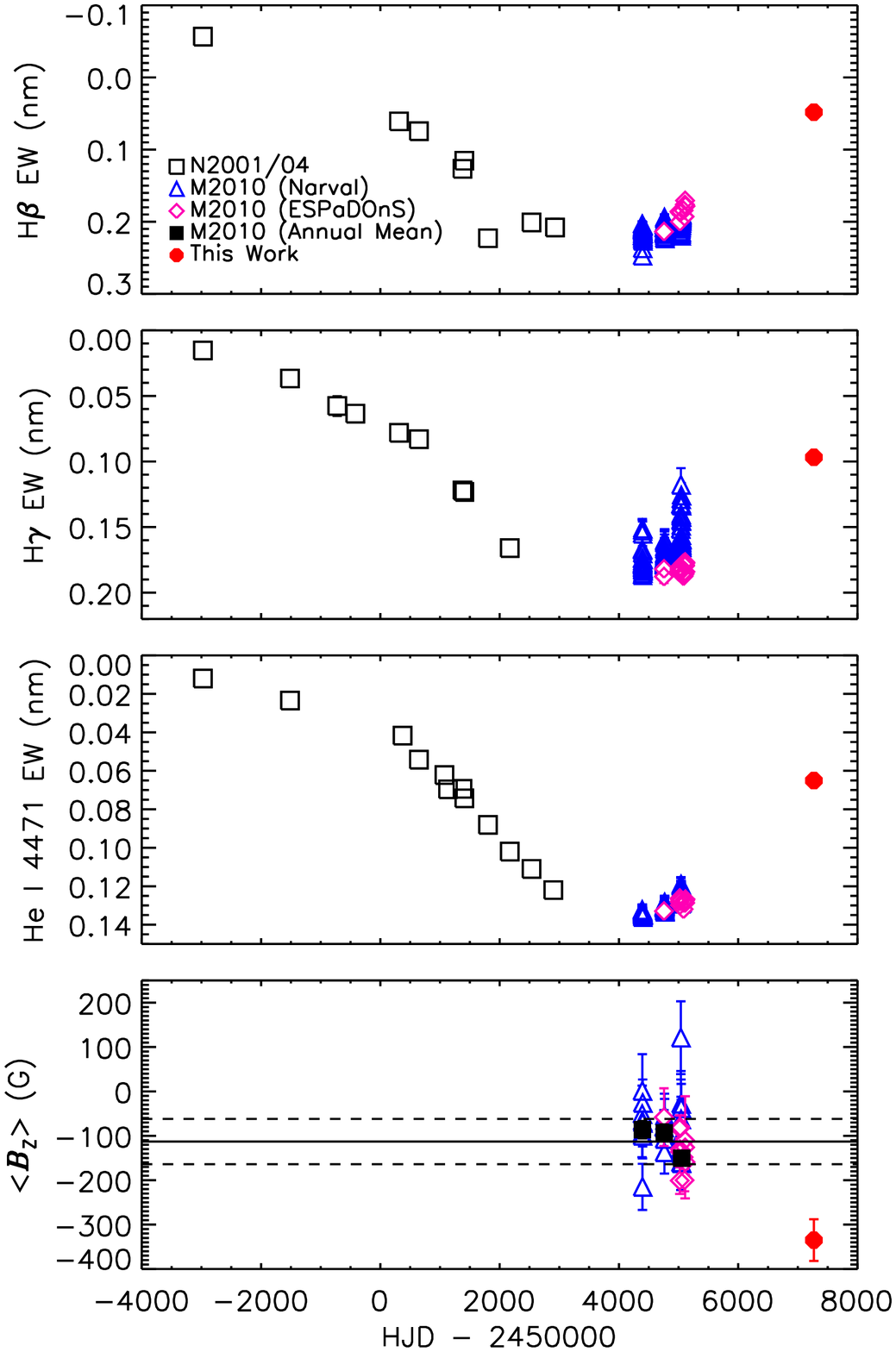}
\caption{EWs and \bz~as functions of time. In the legend, N2001/04 refer to the measurements presented by \protect\cite{2001AA...372..195N} and \protect\cite{2004AA...417..667N}, and M2010 to those of M2010. For the \bz~measurements, measurements from 2007-2009 are shown as nightly means (open symbols) and annual means (filled black squares). The solid and dashed lines show the median \bz~and $\pm$ 1 standard deviation. In 2015 \bz~$=-325 \pm 46$~G, a difference of more than 3$\sigma$ compared with the median \bz~between 2007 and 2009. All measurements were performed using LSD profiles extracted with the line mask used by M2010 (see \S~4). A closer view of the \bz~variation, showing only the epoch spanned by the \bz~measurements, is provided in Fig.\ \ref{bz}.}
\label{ew}
\end{centering}
\end{figure} 

The left column of Fig.\ \ref{var} shows a comparison between the line profiles of H$\gamma$, H$\beta$, and H$\alpha$ in 2015 vs.\ 2007-2009. The Balmer lines are much stronger in the 2015 observation, confirming that HD 108 is moving towards a high state. Comparison of the 2015 observation to the H$\beta$ and H$\gamma$ variability reported by \cite{2001AA...372..195N} and the H$\alpha$ variability reported by M2010 shows that the 2015 data are most similar in appearance to observations collected between 1996 and 1998. As expected, the star's emission lines are not yet at their most intense: the peak strength of H$\beta$, last reported in 1987, was approximately 1.65$\times$ the continuum, as compared to 1.2$\times$ the continuum in the 2015 data. 

The right column of Fig.\ \ref{var} shows the comparison described in the previous paragraph for the He\,{\sc i} 447.1 nm line, the C\,{\sc iii} and N\,{\sc iii} lines near 464 nm, and the He\,{\sc ii} 468.6 nm line. We confirm the same pattern of variability as observed by \cite{2001AA...372..195N}: in comparison to the 2008 data, He\,{\sc ii} and N\,{\sc iii} are essentially unchanged, C\,{\sc iii} is noticeably stronger, and He\,{\sc i} is much weaker, having been significantly filled by emission. 

We measured the following  equivalent widths (EWs) from the 2015 data: for H$\gamma$, we found $0.097\pm0.003$~nm; for H$\beta$, $0.048\pm0.003$~nm; for H$\alpha$, $-0.533\pm0.002$~nm; and for He\,{\sc i} 447.1 nm, $0.065\pm0.002$~nm. While an extended time series of H$\alpha$ EW measurements has not been published, EW time series exist for H$\gamma$, H$\beta$, and He\,{\sc i} 447.1 nm. Fig.\ \ref{ew} shows EW measurements for H$\beta$, H$\gamma$, and He\,{\sc i} 447.1 nm, where we have combined our data with the measurements presented by \cite{2001AA...372..195N}, \cite{2004AA...417..667N}, and M2010. The long-term modulation is apparent in all lines, but is especially clear in He\,{\sc i} as the H lines show a degree of scatter. 

The EW time series confirms the inference from visual comparison of emission lines that the star is in a state similar to that observed in 1998 (HJD~$\sim 2451000$). Assuming the spectroscopic variability to be approximately symmetric about the low state, HD 108 should return to the previously observed maximum emission state in approximately 16 years. This is consistent with the rotation period of $\sim$55 yrs suggested by \cite{2010AA...520A..59N}. If this period is correct, maximum emission should next be observed in 2036. 

\section{Magnetic field diagnosis}

\begin{figure}
\begin{centering}
\includegraphics[width=\hsize]{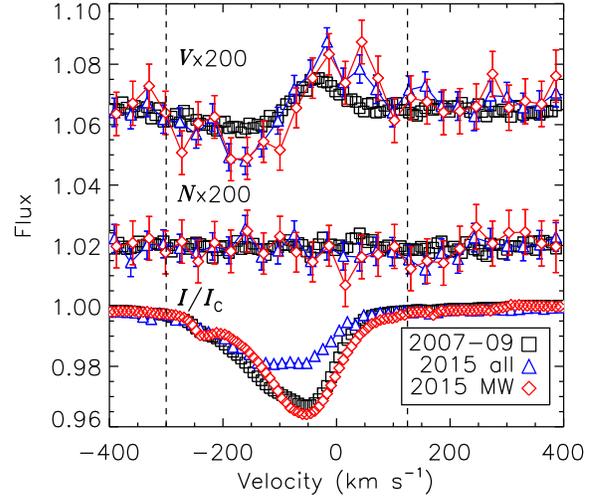}
\caption{LSD profiles. We extracted LSD profiles using the full line list used by M2010 (`all', blue triangles), and a reduced line list consisting of only those lines with the absolute minimum contamination with emission from the stellar wind (`MW', red diamonds). The 2007-2009 grand mean LSD profile is shown for comparison, and was extracted with the full line list. Vertical dotted lines indicate the integration range used for measuring \bz. The Stokes $V$ and $N$ profiles of the 2015 LSD profiles have been re-binned using a 28.8 \kms~velocity pixel.}
\label{lsd}
\end{centering}
\end{figure}

As a first step in evaluating the star's magnetic field we performed Least-Squares Deconvolution (LSD; \citealt{d1997}) using the iLSD package developed by \cite{koch2010}. We used the two line lists published by M2010 (see their Table 2). The first line list contains 17 spectral lines which were manually selected so as to minimize contamination by the wind. The second line list contains the 5 spectral lines identified as having the smallest blue-shifted absorption with respect to the stellar wind, and thus the absolute minimum of contamination by wind emission. In the following we shall refer to the first list as that containing `all' spectral lines (i.e., all those lines included by M2010), and the second line list as the `minimum wind' line mask. The LSD profiles extracted from the 2015 ESPaDOnS observation with these masks are shown in Fig.\ \ref{lsd}, where they are compared to the `low state' grand mean LSD profile obtained by combining all LSD profiles extracted from the observations reported by M2010 using the full line list. 

The amplitude of Stokes $V$ is noticeably stronger in the most recent observation as compared to the 2007-2009 grand mean, however the Stokes $I$ LSD profile extracted using all lines is much weaker. This suggests that many of the spectral lines included in the larger line mask are in fact significantly affected by the stellar magnetosphere, notwithstanding the attempt by M2010 to select lines with only small contamination with wind emission. Conversely, the Stokes $I$ LSD profile extracted using the minimum wind mask is similar in depth to that of the 2007-2009 `low state' grand mean, indicating that this smaller line mask is largely successful in eliminating wind contamination. The 2015 Stokes $V$ profiles extracted with the two masks are similar, although the minimum-wind mask yields a lower SNR due to the smaller number of included lines. This supports the assumption that Stokes $V$ is unaffected by circumstellar emission, as expected given that the magnetic field should be much stronger at the photosphere than in the circumstellar environment, and confirming the results of M2010 who performed the same comparison. 

\begin{figure}
\begin{centering}
\includegraphics[width=\hsize]{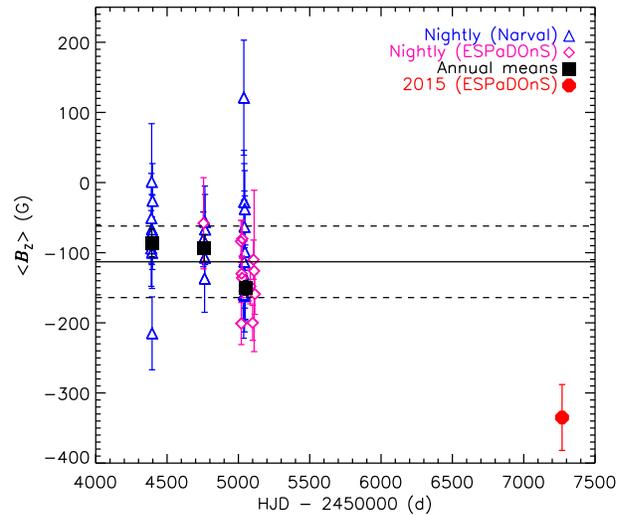}
\caption{\bz~as a function of time, as in the bottom panel of Fig.\ \ref{ew} but zoomed in to show the epoch spanned by the \bz~measurements.}
\label{bz}
\end{centering}
\end{figure}

To evaluate the longitudinal magnetic field \bz~(e.g.\ \citealt{mat1989}), we would ideally like to measure \bz~corresponding as closely as possible to the true photospheric value, thus as far as possible contamination from magnetospheric emission should be avoided. This can be done simply by using the minimum-wind profile, but this sacrifices precision in Stokes $V$. Instead, we used the LSD profiles extracted using the full line mask, but fixed the Stokes $I$ EW to the maximum EW measured in this dataset. This makes the assumption that all variability in Stokes $I$ is a consequence of the magnetosphere, and that the maximum EW gives the best approximation of the true photospheric line strength. This assumption seems warranted given the much lower level of variability in the minimum wind LSD profiles on either short or long timescales. 

These \bz~measurements are shown as a function of time in the bottom panel of Fig.\ \ref{ew}, where they are compared to the EWs, and alone in Fig.\ \ref{bz} where the time axis is zoomed in to show only the epoch spanned by the magnetic data. The weighted mean \bz~measurement in the 2007-2009 epoch is -128$\pm$8 G, with a standard deviation for individual measurements of 46 G (solid and dashed lines in Figs.\ \ref{ew} and \ref{bz}), close to the mean error bar of 54 G. \bz~in a given year is thus consistent with no variation. The annual weighted mean \bz~(black squares in Figs.\ \ref{ew} and \ref{bz}) are suggestive of a slight increase in the strength of \bz~over time. These results are consistent with those of M2010, confirming that in the 2007-2009 epoch the wind was minimally affecting the lines used for LSD. 


In contrast, in 2015 \bz$=-325\pm46$~G, approximately 3 times the strength measured in 2007-2009. If instead we use the minimum wind LSD Stokes $V$ profile to evaluate \bz, \bz$=-375 \pm 93$ G, which is consistent within error bars. From the LSD Stokes $I$ and $V$ profiles extracted with the full line mask used by M2010, we obtain \bz~$=-643 \pm 107$~G, where the much higher \bz~is a result of the smaller Stokes $I$ EW. 

\section{Magnetic, Magnetospheric, and Rotational Parameters}

\begin{table}
\centering
\caption[Stellar and Magnetic parameters]{Stellar, magnetic, and rotational parameters}
\begin{tabular}{lrl}
\hline
\hline
Parameter & Quantity & Origin \\
\\
\hline
\multicolumn{3}{c}{Stellar parameters} \\
$\log{(L_{\rm bol}/L_\odot)}$ & 5.7$\pm$0.1 & M2010 \\
$T_{\rm eff}$ (kK) & 35$\pm$2 & M2010 \\
$R_*$ ($R_\odot$) & 19.4$\pm$1.5 & M2010 \\
$M_*$ ($M_\odot$) &  42$\pm$5 & This work \\
$M_{\rm ZAMS}$ ($M_\odot$) & 50$\pm$3 & This work \\
Age (Myr) & 3.3$\pm$0.3 & This work \\
$\log{[\dot{M}/(M_\odot {\rm yr}^{-1})]}$ & -5.55$\pm$0.17 & This work \\
$v_\infty$ (\kms) & 2000$\pm$300 & \cite{2012MNRAS.422.2314M} \\
\\
\multicolumn{3}{c}{Rotational parameters} \\
$P_{\rm rot}$ (yr) & 55 & \cite{2010AA...520A..59N} \\
$W$ & (8$\pm$1)$\times 10^{-5}$ & This work \\
$R_{\rm K}$ ($R_*$) & 560$\pm$70 & This work \\
\\
\multicolumn{3}{c}{Magnetic parameters} \\
\bz$_{\rm max}$ (G) & -325$\pm$45 & This work \\
$B_{\rm d}$ (G) & $\ge$1150 & This work \\
$\eta_*$ & $\ge$11.5 & This work \\
$R_{\rm A}$ ($R_*$) & $\ge$2.1 & This work \\
$\tau_{\rm J}$ (Myr) & $\le$0.5 & This work \\
$t_{\rm S,max}$ (Myr) & $\le$5 & This work \\
$\log{L_{\rm X}}_{\rm obs}$ (erg s$^{-1}$) & 33.0 & \cite{2014ApJS..215...10N} \\
$\log{L_{\rm X, XADM}}$ (erg s$^{-1}$) & 33.3$\pm$0.4 & This work \\
\hline
\hline
\end{tabular}
\label{star_prop}
\end{table}

\begin{figure}
\begin{centering}
\includegraphics[width=\hsize]{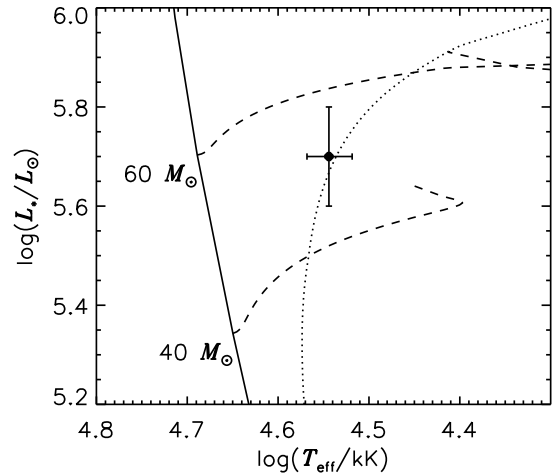}
\caption{HD 108's position on the HRD. The dashed lines indicate evolutionary tracks for 40 and 60 $M_\odot$ models; the solid line indicates the ZAMS; and the dotted line indicates the $\log{(t/{\rm yr})}=6.5$ isochrone. Evolutionary tracks and isochrones are from the non-rotating models presented by \protect\cite{ekstrom2012}.}
\label{hrd}
\end{centering}
\end{figure} 

HD 108's stellar, magnetic, magnetospheric, and rotational parameters are summarized in Table \ref{star_prop}. The theoretical framework concerning the rotational and magnetic characteristics of stellar magnetospheres has been summarized by \cite{petit2013}, whose development we follow to determine magnetospheric confinement radii, rotation parameters, and spindown timescales.

The lower limit to the dipole magnetic field strength $B_{\rm d}$ can be inferred from the maximum \bz~measurement \bz$_{\rm max}$ and the limb darkening coefficient \citep{preston1967}. According to the tables calculated by \cite{diazcordoves1995}, a star with HD 108's \teff~and $\log{g}$ should have a limb darkening coefficient of $\sim$0.3. Using the formula provided by \cite{preston1967}, $B_{\rm d} \ge 3.5 |\langle B_{\rm Z} \rangle |_{\rm max} = 1150 \pm 160$~G. 

The extent of the star's magnetically confined wind is given by the Alfv\'en radius $R_{\rm A}$ \citep{ud2002}. This is determined via a scaling relation with the wind magnetic confinement parameter $\eta_*$, which is the ratio of magnetic to kinetic energy density in the stellar wind \citep{ud2002}. In order to evaluate $\eta_*$ and $R_{\rm A}$, we must know the star's wind parameters, for which we must first determine the stellar parameters. 

We adopt \teff$=35\pm2$~kK and $\log{(L/L_\odot)} = 5.7 \pm 0.1$, as given by M2010 based on a spectroscopic analysis of archival IUE observations. The stellar radius is then $R_* = 19.4\pm1.5 R_\odot$. Placing the star on the HRD (Fig.\ \ref{hrd}) and comparing to the non-rotating evolutionary tracks and isochrones calculated by \cite{ekstrom2012}, we find the star to have a zero-age main sequence (ZAMS) mass of $M_{\rm ZAMS} = 50 \pm 3 M_\odot$, a present mass of $M_* = 42 \pm 5 M_\odot$, and an age of $3.3\pm0.3$ Myr. These are similar to the parameters determined by M2010, $M_*=43~M_\odot$ and $t=4 \pm 1$~Myr, where their slightly greater age is due to their adoption of the rotating evolutionary tracks published by \cite{2003AA...411..543M}. 

We use the mass loss recipe of \cite{vink2001}, which yields a surface mass flux of $\log{[\dot{M}/(M_\odot~{\rm yr}^{-1})]} = -5.55 \pm 0.17$, calculated using a wind terminal velocity \vinf$=2000 \pm 300$ \kms~as measured from UV lines by \cite{2012MNRAS.422.2314M} (this \vinf~is compatible with the value determined from scaling the star's escape velocity by 2.6, as suggested by \citeauthor{vink2001}). This mass-loss rate is much higher than that determined via spectral modelling of ultraviolet resonance lines by \cite{2012MNRAS.422.2314M}, $\log{\dot{M}} = -7 \pm 1$. However, \citeauthor{2012MNRAS.422.2314M} used spherically symmetric mass-loss models which they noted were unable to simultaneously reproduce emission and absorption features. Comparisons of spherically symmetric to magnetohydrodyanmic fits to the UV lines of HD 57682 \citep{2012MNRAS.426.2208G} and HD 191612 \citep{2013MNRAS.431.2253M} have also found that MHD models require mass-loss rates about 1 dex higher than those measured using spherically symmetric models. MHD mass-loss rates are furthermore similar to those calculated by the \cite{vink2001} recipe. We therefore adopt the \citeauthor{vink2001} mass-loss rate. From Eqns.\ 7 and 8 of \cite{ud2002}, we then find $\eta_* \ge 11.5$ and $R_{\rm A} \ge 2.1~R_*$.

The corotation or Kepler radius $R_{\rm K}$ is determined via the rotation parameter $W \equiv v_{\rm eq} / v_{\rm orb}$, i.e.\ the ratio of the equatorial velocity to the orbital velocity \citep{ud2008}. Assuming $P_{\rm rot}=55$~yr, we find $W = (8 \pm 1)\times 10^{-5}$ and $R_{\rm K} = 560 \pm 70~R_*$ (\citealt{ud2008}, Eqns. 11 and 14). 

\cite{ud2009} provided a scaling relation for the spindown timescale $\tau_{\rm J}$ due to angular momentum loss via the magnetosphere. This scales with the star's moment of inertia, the (non-magnetic) mass-loss timescale, and $R_{\rm A}$. Assuming initially critical rotation, i.e.\ $W_0 = 1$, the maximum spindown time $t_{\rm S,max}$ (i.e.\ the time required for the star to have decelerated from critical rotation to its current rotation rate) can be estimated from $\tau_{\rm J}$ and $W$. Using $\dot{M}$, $R_{\rm A}$, and $W$ as determined above, we find $\tau_{\rm J} \le 0.5$ Myr (\citealt{petit2013}, Eqn. 12), and $t_{\rm S, max} \le 5$ Myr. This is somewhat higher than the age inferred from the HRD, $t=3.3 \pm 0.3$~Myr, but is closer than the 8~Myr spindown age found by \cite{petit2013}. If instead the mass-loss rate measured from UV lines by \cite{2012MNRAS.422.2314M} is used, $t_{\rm S, max} < 25$~Myr, much longer than the main sequence lifetime of the star. 

HD 108 has an X-ray luminosity of $\log{[L_{\rm X} / ({\rm erg~s^{-1}})]} = 33$ \citep{2014ApJS..215...10N}, which is overluminous in comparison to similar non-magnetic stars. \cite{ud2014} developed a semi-analytic scaling relationship that has proven successful in predicting the X-ray luminosities of most magnetic OB stars, although HD 108 was predicted to be about 0.5 dex less luminous than observed \citep{2014ApJS..215...10N}. Using the lower limit on $B_{\rm d}$ determined from the new magnetic data, we find that the XADM model predicts $\log{[L_{\rm X} / ({\rm erg~s^{-1}})]} \ge 33.3 \pm 0.4$, consistent with the observed X-ray luminosity. The uncertainty accounts for the uncertainty in \vinf, which strongly affects $\log{L_{\rm X}}$, and the uncertain efficiency of X-ray production, which may be between 5\% and 20\% depending on the degree of self-absorption of shock-produced X-rays within the magnetosphere's cool plasma. 

\section{Discussion}

The wind-sensitive lines of HD 108 show variability on two timescales: a long-term modulation, and short-term variability manifesting as scatter in a given epoch. Similar scatter has been observed in the H$\alpha$ EWs of the magnetic O stars $\theta^1$ Ori C \citep{2008AA...487..323S}, HD 148937 \citep{wade2012a}, and CPD~$-28^\circ 2561$ \citep{2015MNRAS.447.2551W}. This scatter can be qualitatively reproduced in 3D magnetohydrodynamic simulations as a consequence of time-variable structure within the magnetosphere \citep{ud2013}. A period search by M2010 on the EWs of variable spectral lines found no stable periods, suggesting stochastic behaviour within the magnetosphere. We performed our own analysis of H$\alpha$ EW data and have confirmed their conclusion. 

The long-term modulation of magnetic O-type star EWs seen in wind sensitive lines is produced by the changing projection of the stellar magnetosphere on the sky. If the rotational and magnetic axes are misaligned, then as the star rotates the angle between the line of sight and the magnetic axis changes. Magnetically confined plasma collects in a disk or torus-like structure in closed loops surrounding the magnetic equator, and corotates with the star. Thus, as the star rotates, the magnetosphere is seen from varying perspectives. When the magnetosphere is closest to face-on, emission strength is at a maximum, whereas when it is seen edge-on, emission is at a minimum. These phases correspond to the magnetic axis being, respectively, closest to parallel or perpendicular to the line of sight, thus also corresponding to maximum and minimum \bz. 

\begin{figure}
\begin{centering}
\includegraphics[width=\hsize]{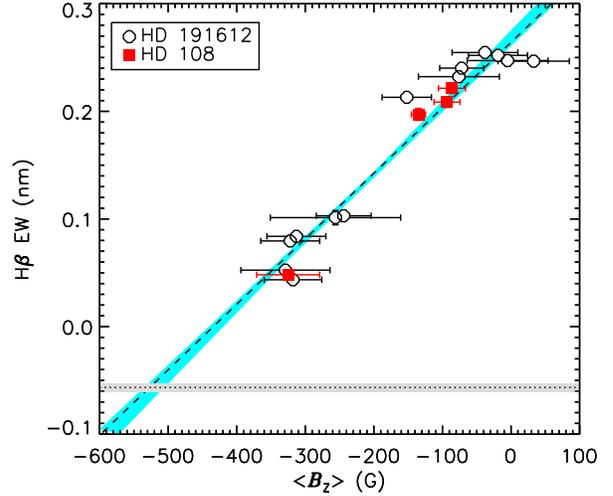}
\caption{Extrapolation of HD 108's maximum \bz~using H$\beta$ EWs. The dashed line shows a linear regression of HD 191612's \bz~and H$\beta$ EW measurements (open circles). The blue shaded region indicates the uncertainty in the regression, derived via iterative removal of individual datapoints. HD 108's measurements are shown by solid red squares. The dotted line shows HD 108's H$\beta$ EW at emission maximum, and the grey shaded region indicates the mean uncertainty in the EW time series (no uncertainty was provided for the original measurement). This method estimates \bz$_{\rm max} \sim 550$ G.}
\label{bz_extrap}
\end{centering}
\end{figure} 

The magnetic data presented in \S~4 indicate that HD 108's surface magnetic dipole is at least twice as strong as the previously reported lower limit. However, since the star has not yet returned to emission maximum, and since \bz~and emission line EWs tend to correlate, it is likely that \bz~will continue to increase. If this is the case, to estimate the maximum strength of \bz, we used the \bz~and H$\beta$ EWs for HD 191612 (O6-8f?p), which has the most similar stellar parameters to HD 108's of any other magnetic O-type star (\teff~$=36\pm1$ kK, $\log{L}=5.4 \pm 0.2$), \bz~of similiar magnitude, and complete phase coverage of both \bz~and H$\beta$ \citep{wade2011}. We used H$\beta$ rather than the more sensitive H$\alpha$ lines as HD 108's H$\alpha$ time series does not extend to phases of emission maximum. Fig.\ \ref{bz_extrap} shows a linear regression of H$\beta$ EW vs.\ \bz. The correlation coefficient for HD 191612 is $r^2 = 0.96$, indicating a good correlation. HD 108's measurements fall along this regression line, suggesting that using the regression to extrapolate \bz$_{\rm max}$ is not unreasonable. The dotted line shows HD 108's H$\beta$ EW at emission maximum; it intersects the regression at \bz~$\sim-550$~G, implying that $B_{\rm d} > 1.9$~kG. 

While the distribution of surface magnetic dipole strengths amongst magnetic O-type stars ranges from a few hundred G to 20 kG, the majority of such stars have 1~kG~$<~B_{\rm d}~<$~4~kG \citep{petit2013,2015ASPC..494...30W}. The 2~kG lower limit obtained from the HD 191612 extrapolation is very close to the centre of the distribution, while the lower limit determined from the 2015 \bz~measurement is within 1 standard deviation. This suggests that HD 108 is unlikely to have a magnetic field stronger than about 4 kG.

Recalculating the magnetospheric parameters and spindown timescales with the lower limit inferred from HD 191612's H$\beta$ EWs as the magnetic field strength of HD 108 yields $\eta_* > 33$, $R_{\rm A} > 2.7 R_*$, $\tau_{\rm J} \le 0.3$~Myr and $t_{\rm S,max} \le 3$~Myr. The upper limit on the maximum spindown age is approximately the same as the age inferred from the non-rotating evolutionary models of \cite{ekstrom2012}, $t = 3.3 \pm 0.3$~Myr (Fig.\ \ref{hrd}). The X-ray luminosity predicted by the XADM model, assuming 10\% efficiency, increases to $\log{L_{\rm X}} \ge 33.6$, about 0.6 dex higher than the observed X-ray luminosity, and outside the 0.4 dex uncertainty. This difference is small enough that it can be reconciled by reducing the efficiency to 2\%, and/or if the X-ray luminosity is rotationally modulated, as has been observed for some Of?p stars, e.g.\ HD 191612 for which $\log{L_{\rm X}}$ varies by about 0.13 dex throughout a rotational cycle \citep{2014ApJS..215...10N}. Efficiency is expected to decrease with stronger magnetic confinement, due to the higher density and greater volume of the circumstellar plasma, which absorbs a greater fraction of X-rays \citep{2014ApJS..215...10N}. HD 108 has stronger emission than any magnetic O-type star but NGC 1624-2, which has by far both the strongest magnetic field and the strongest magnetospheric emission of any magnetic O-type star \citep{2013MNRAS.428.1686G}, and is also the most X-ray underluminous with respect to the XADM model \citep{2014ApJS..215...10N}. Given this, greater absorption of X-rays by HD 108's magnetosphere than in those of most other magnetic O-type stars would make sense. 

\begin{figure}
\begin{centering}
\includegraphics[width=\hsize]{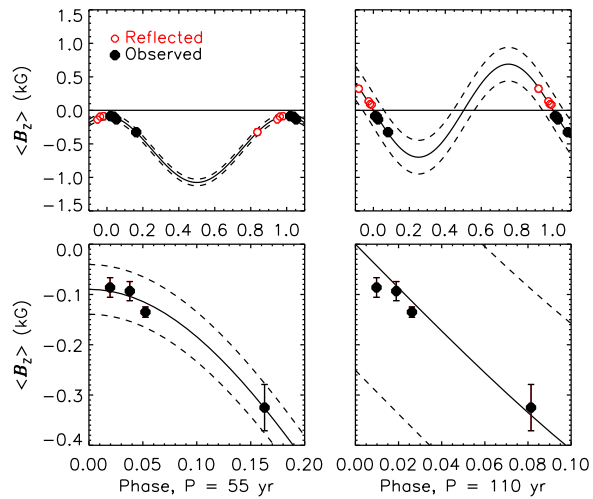}
\caption{Extrapolation of HD 108's \bz~curve using either a 55~yr period (left) or a 110~yr period (right). Filled circles correspond to annual mean \bz~measurements, open circles indicate the estimated \bz~obtained from reflecting \bz~about phase 0. In the case of a 110-yr period, since the EW variation must be a double wave, both magnetic poles must be visible, thus the polarity of the reflected measurements must be reversed. Reflected measurements were used to constrain least-squares sinusoidal fits, indicated by solid curves; dashed curves show the 1$\sigma$ uncertainty in the fits. The bottom panels show the fit in the vicinity of the observed and reflected \bz~measurements.}
\label{bz_crv}
\end{centering}
\end{figure} 

If only one magnetic pole is visible throughout a rotational cycle, the H$\alpha$ EW will show a single-wave variation, with a single emission maximum at \bz$_{\rm max}$ and a single minimum at \bz~$\sim 0$ (e.g.\ HD 191612, and NGC 1624-2; \citealt{wade2011, wade2012b}). If both magnetic poles are visible, the emission strength will show a double-wave variation, with two local maxima corresponding to the positive and negative extrema of the \bz~curve, and two local minima corresponding to \bz~$=0$ (e.g.\ $\theta^1$ Ori C, HD 57682, HD 148937, and CPD~$-28^\circ 2561$; \citealt{2008AA...487..323S, 2012MNRAS.426.2208G, wade2012a, 2015MNRAS.447.2551W}). While HD 108's variability is consistent with a 55 year period, with no more than about 50\% of a rotation cycle covered, a double wave variation cannot be ruled out on the basis of spectroscopic data alone, in which case the rotation period would be 110 years. 

It may be significant that all \bz~measurements to date have been negative. For a double-wave variation, emission minimum corresponds to a polarity change in \bz: thus, if the most recent data had been of positive polarity, the period would have to be 110 years. To explore this further, we calculated least-squares sinusoidal fits to \bz~using both periods. These are shown in Fig.\ \ref{bz_crv}. To help constrain the fits, we estimated \bz~under the assumption that a sinusoidal \bz~curve must be symmetrical about phase 0, which we define here using ${\rm JD0} = 2454000$, corresponding to the time of minimum emission (see Fig.\ \ref{ew}). For a 55 year period, \bz~should be negative at all phases. For a 110~yr period, \bz~should be positive between phases 0.5 and 1.0, thus, the polarity of the reflected \bz~estimates was reversed. 

In the case of a 110~yr period, there should have been an approximately linear decrease in \bz~between 2007 and 2015, whereas for a 55~yr period the 2007-2009 epoch should have corresponded to an extremum of the \bz~curve, with the rate of change increasing from that epoch to the present. The latter case seems to be a better match to the observations, especially considering the higher precision of the 2007-2009 annual mean measurements as compared to the 2015 data. This is reflected in the smaller uncertainty in the fit obtained using a 55~yr period.

\section{Conclusions}

HD 108 is moving back into a high emission state. The line profiles are similar to those seen in 1998. This is consistent with the 55 year rotation period suggested by \cite{2010AA...520A..59N}.

The new magnetic measurement shows that HD 108's magnetic field is at least twice as strong ($B_{\rm d} > 1.2$~kG) as the previous lower limit. This increase in \bz~accompanying the increase in emission strength confirms the oblique rotator model as a unified explanation for HD 108's magnetic and spectroscopic variability. Comparison of HD 108's H$\beta$ EW curve to those of the similar magnetic O-type star HD 191612 suggests that $B_{\rm d}$ is $>2$~kG. This places HD 108 within one standard deviation of the centre of the observed $B_{\rm d}$ distribution of magnetic O-type stars. A 2 kG dipole yields better agreement between the spindown timescale and the stellar age inferred from HD 108's position on the HRD. 

The higher lower limit to $B_{\rm d}$ also resolves the discrepancy between HD 108's observed X-ray luminosity and that predicted by the XADM model: HD 108 is now slightly less luminous than predicted by XADM, as is the case for all other magnetic O-type stars. Indeed, bringing the observed and predicted X-ray luminosities into agreement now requires an efficiency of $\sim$2\%, somewhat less than the 5-10\% required for most other magnetic O-type stars. This reduced efficiency may be consistent with the fact that HD 108's H emission is stronger than any star but NGC 1624-2, implying a larger magnetosphere that absorbs a greater fraction of the X-rays produced by the magnetically confined wind shocks.

The increased rate of change in \bz~between the 2007-2009 epoch, when \bz~was essentially flat, and the 2015 measurement, is more consistent with a single-wave EW variation in which only one magnetic pole is visible. This indicates that the period is likely 55 rather than 110 years. 

Further magnetic data will be essential to constraining the star's surface magnetic field strength. From the EW variation, and assuming a 55 year rotation period, magnetic maximum should next occur in 2036. Until then, a new magnetic measurement should be collected at least once every 5 years, in order to sample the rotational phase curve in increments of at least 0.1. Additional X-ray observations should also be obtained in similar intervals, in order to determine to what degree rotational modulation can explain the discrepancy between observed and predicted X-ray luminosities. 

Future spectroscopic data may also be instrumental in distinguishing between 55~yr and 110~yr periods. Unless $i+\beta$ is exactly $180^\circ$, the EW curve will not be perfectly symmetric between times of positive and negative magnetic polarity. Thus, if the next emission maximum (which will correspond to the next magnetic maximum) is substantially stronger or weaker than the previous, this will be good evidence that the period is actually 110~yr, while if the maxima are of the same strength, it will be more likely that $P_{\rm rot}=55$~yr. Comparison of observed EW curves to those predicted by MHD models of HD 108's stellar wind and magnetosphere, as performed for HD 57682 \citep{2012MNRAS.426.2208G}, HD 191612 \citep{sund2012}, and CPD $-28^\circ 2561$ \citep{2015MNRAS.447.2551W}, will be helpful in determining the star's magnetic geometry, as such models can help to constrain the inclination angle. Due to its extremely slow rotation HD 108 also presents an excellent target for spectral monitoring, which could be used to explore the characteristic timescales of turbulent plasma flows within stellar magnetospheres.

The conclusion that HD 108's spindown age and stellar age are compatible is tentative, as it relies upon evolutionary models that do not account for the effects of magnetic fields on stellar structure. \cite{meynet2011} explored the impact of rotational braking and the inhibition of mixing on stellar evolution, however a truly self-consistent treatment has yet to be performed. Future models should investigate the inter-relation of mass-loss quenching due to magnetic wind confinement, as well as the decline in the surface magnetic field strength due either to flux conservation in an expanding stellar atmosphere, and/or to magnetic flux decay \citep{2008AA...481..465L,2016A&A...592A..84F}. These effects have the potential to modify stellar evolution directly, while at the same time, the evolution of the star may have an influence on the magnetic field, and hence on the magnetosphere and magnetic braking. When such models are available, the gyrochronological ages of massive, magnetic stars should be revisited. 
\\
\\
{\small {\em Acknowledgements} This work has made use of the VALD database, operated at Uppsala University, the Institute of Astronomy RAS in Moscow, and the University of Vienna. MS and GAW acknowledge support from the Natural Science and Engineering Research Council of Canada (NSERC). We acknowledge the Canadian Astronomy Data Centre (CADC).

\bibliography{bib_dat.bib}

\begin{thebibliography}{42}
\expandafter\ifx\csname natexlab\endcsname\relax\def\natexlab#1{#1}\fi

\bibitem[{{Babel} \& {Montmerle}(1997)}]{bm1997}
{Babel} J., {Montmerle} T., 1997, \apjl, 485, L29

\bibitem[{{Barannikov}(2007)}]{2007IBVS.5756....1B}
{Barannikov} A.~A., 2007, Information Bulletin on Variable Stars, 5756, 1

\bibitem[{{Diaz-Cordoves}, {Claret} \& {Gimenez}(1995){Diaz-Cordoves},
  {Claret}, \& {Gimenez}}]{diazcordoves1995}
{Diaz-Cordoves} J., {Claret} A., {Gimenez} A., 1995, \aaps, 110, 329

\bibitem[{{Donati} {et~al}\mbox{.}(1997){Donati}, {Semel}, {Carter}, {Rees}, \&
  {Collier Cameron}}]{d1997}
{Donati} J.-F., {Semel} M., {Carter} B.~D., {Rees} D.~E., {Collier Cameron} A.,
  1997, MNRAS, 291, 658

\bibitem[{{Ekstr{\"o}m} {et~al}\mbox{.}(2012){Ekstr{\"o}m}, {Georgy},
  {Eggenberger}, {Meynet}, {Mowlavi}, {Wyttenbach}, {Granada}, {Decressin},
  {Hirschi}, {Frischknecht}, {Charbonnel}, \& {Maeder}}]{ekstrom2012}
{Ekstr{\"o}m} S. {et~al.}, 2012, \aap, 537, A146

\bibitem[{{Fossati} {et~al}\mbox{.}(2016){Fossati}, {Schneider}, {Castro},
  {Langer}, {Sim{\'o}n-D{\'{\i}}az}, {M{\"u}ller}, {de Koter}, {Morel},
  {Petit}, {Sana}, \& {Wade}}]{2016A&A...592A..84F}
{Fossati} L. {et~al.}, 2016, \aap, 592, A84

\bibitem[{{Grunhut} {et~al}\mbox{.}(2013){Grunhut}, {Wade}, {Leutenegger},
  {Petit}, {Rauw}, {Neiner}, {Martins}, {Cohen}, {Gagn{\'e}}, {Ignace},
  {Mathis}, {de Mink}, {Moffat}, {Owocki}, {Shultz}, {Sundqvist}, \& {MiMeS
  Collaboration}}]{2013MNRAS.428.1686G}
{Grunhut} J.~H. {et~al.}, 2013, \mnras, 428, 1686

\bibitem[{{Grunhut} {et~al}\mbox{.}(2012){Grunhut}, {Wade}, {Sundqvist},
  {ud-Doula}, {Neiner}, {Ignace}, {Marcolino}, {Rivinius}, {Fullerton},
  {Kaper}, {Mauclaire}, {Buil}, {Garrel}, {Ribeiro}, \&
  {Ubaud}}]{2012MNRAS.426.2208G}
{Grunhut} J.~H. {et~al.}, 2012, \mnras, 426, 2208

\bibitem[{{Kochukhov}, {Makaganiuk} \& {Piskunov}(2010){Kochukhov},
  {Makaganiuk}, \& {Piskunov}}]{koch2010}
{Kochukhov} O., {Makaganiuk} V., {Piskunov} N., 2010, \aap, 524, A5

\bibitem[{{Landstreet} {et~al}\mbox{.}(2008){Landstreet}, {Silaj}, {Andretta},
  {Bagnulo}, {Berdyugina}, {Donati}, {Fossati}, {Petit}, {Silvester}, \&
  {Wade}}]{2008AA...481..465L}
{Landstreet} J.~D. {et~al.}, 2008, \aap, 481, 465

\bibitem[{{Maeder} \& {Meynet}(2003)}]{2003AA...411..543M}
{Maeder} A., {Meynet} G., 2003, \aap, 411, 543

\bibitem[{{Marcolino} {et~al}\mbox{.}(2013){Marcolino}, {Bouret}, {Sundqvist},
  {Walborn}, {Fullerton}, {Howarth}, {Wade}, \&
  {ud-Doula}}]{2013MNRAS.431.2253M}
{Marcolino} W.~L.~F., {Bouret} J.-C., {Sundqvist} J.~O., {Walborn} N.~R.,
  {Fullerton} A.~W., {Howarth} I.~D., {Wade} G.~A., {ud-Doula} A., 2013,
  \mnras, 431, 2253

\bibitem[{{Marcolino} {et~al}\mbox{.}(2012){Marcolino}, {Bouret}, {Walborn},
  {Howarth}, {Naz{\'e}}, {Fullerton}, {Wade}, {Hillier}, \&
  {Herrero}}]{2012MNRAS.422.2314M}
{Marcolino} W.~L.~F. {et~al.}, 2012, \mnras, 422, 2314

\bibitem[{{Martins} {et~al}\mbox{.}(2010){Martins}, {Donati}, {Marcolino},
  {Bouret}, {Wade}, {Escolano}, {Howarth}, \& {Mimes
  Collaboration}}]{2010MNRAS.407.1423M}
{Martins} F., {Donati} J.-F., {Marcolino} W.~L.~F., {Bouret} J.-C., {Wade}
  G.~A., {Escolano} C., {Howarth} I.~D., {Mimes Collaboration}, 2010, \mnras,
  407, 1423

\bibitem[{{Mathys}(1989)}]{mat1989}
{Mathys} G., 1989, FCPh, 13, 143

\bibitem[{{Meynet}, {Eggenberger} \& {Maeder}(2011){Meynet}, {Eggenberger}, \&
  {Maeder}}]{meynet2011}
{Meynet} G., {Eggenberger} P., {Maeder} A., 2011, \aap, 525, L11

\bibitem[{{Mikul{\'a}{\v s}ek} {et~al}\mbox{.}(2011){Mikul{\'a}{\v s}ek},
  {Krti{\v c}ka}, {Henry}, {Jan{\'{\i}}k}, {Zverko}, {{\v Z}i{\v z}{\v
  n}ovsk{\'y}}, {Zejda}, {Li{\v s}ka}, {Zv{\v e}{\v r}ina}, {Kudrjavtsev},
  {Romanyuk}, {Sokolov}, {L{\"u}ftinger}, {Trigilio}, {Neiner}, \& {de
  Villiers}}]{miku2011}
{Mikul{\'a}{\v s}ek} Z. {et~al.}, 2011, \aap, 534, L5

\bibitem[{{Mikul{\'a}{\v s}ek} {et~al}\mbox{.}(2008){Mikul{\'a}{\v s}ek},
  {Krti{\v c}ka}, {Henry}, {Zverko}, {{\v Z}i{\v z}{\aa}ovsk{\'y}},
  {Bohlender}, {Romanyuk}, {Jan{\'{\i}}k}, {Bo{\v z}i{\'c}}, {Kor{\v
  c}{\'a}kov{\'a}}, {Zejda}, {Iliev}, {{\v S}koda}, {{\v S}lechta}, {Gr{\'a}f},
  {Netolick{\'y}}, \& {Ceniga}}]{miku2008}
{Mikul{\'a}{\v s}ek} Z. {et~al.}, 2008, \aap, 485, 585

\bibitem[{{Naz{\'e}} {et~al}\mbox{.}(2014){Naz{\'e}}, {Petit}, {Rinbrand},
  {Cohen}, {Owocki}, {ud-Doula}, \& {Wade}}]{2014ApJS..215...10N}
{Naz{\'e}} Y., {Petit} V., {Rinbrand} M., {Cohen} D., {Owocki} S., {ud-Doula}
  A., {Wade} G.~A., 2014, \apjs, 215, 10

\bibitem[{{Naz{\'e}} {et~al}\mbox{.}(2004){Naz{\'e}}, {Rauw}, {Vreux}, \& {De
  Becker}}]{2004AA...417..667N}
{Naz{\'e}} Y., {Rauw} G., {Vreux} J.-M., {De Becker} M., 2004, \aap, 417, 667

\bibitem[{{Naz{\'e}} {et~al}\mbox{.}(2010){Naz{\'e}}, {Ud-Doula}, {Spano},
  {Rauw}, {De Becker}, \& {Walborn}}]{2010AA...520A..59N}
{Naz{\'e}} Y., {Ud-Doula} A., {Spano} M., {Rauw} G., {De Becker} M., {Walborn}
  N.~R., 2010, \aap, 520, A59

\bibitem[{{Naz{\'e}}, {Vreux} \& {Rauw}(2001){Naz{\'e}}, {Vreux}, \&
  {Rauw}}]{2001AA...372..195N}
{Naz{\'e}} Y., {Vreux} J.-M., {Rauw} G., 2001, \aap, 372, 195

\bibitem[{{Petit} {et~al}\mbox{.}(2013){Petit}, {Owocki}, {Wade}, {Cohen},
  {Sundqvist}, {Gagn{\'e}}, {Ma{\'{\i}}z Apell{\'a}niz}, {Oksala}, {Bohlender},
  {Rivinius}, {Henrichs}, {Alecian}, {Townsend}, {ud-Doula}, \& {MiMeS
  Collaboration}}]{petit2013}
{Petit} V. {et~al.}, 2013, \mnras, 429, 398

\bibitem[{{Preston}(1967)}]{preston1967}
{Preston} G.~W., 1967, \apj, 150, 547

\bibitem[{{Pyper} {et~al}\mbox{.}(1998){Pyper}, {Ryabchikova}, {Malanushenko},
  {Kuschnig}, {Plachinda}, \& {Savanov}}]{1998A&A...339..822P}
{Pyper} D.~M., {Ryabchikova} T., {Malanushenko} V., {Kuschnig} R., {Plachinda}
  S., {Savanov} I., 1998, \aap, 339, 822

\bibitem[{{Stahl} {et~al}\mbox{.}(2008){Stahl}, {Wade}, {Petit}, {Stober}, \&
  {Schanne}}]{2008AA...487..323S}
{Stahl} O., {Wade} G., {Petit} V., {Stober} B., {Schanne} L., 2008, \aap, 487,
  323

\bibitem[{{Sundqvist} {et~al}\mbox{.}(2012){Sundqvist}, {ud-Doula}, {Owocki},
  {Townsend}, {Howarth}, \& {Wade}}]{sund2012}
{Sundqvist} J.~O., {ud-Doula} A., {Owocki} S.~P., {Townsend} R.~H.~D.,
  {Howarth} I.~D., {Wade} G.~A., 2012, MNRAS, 423, L21

\bibitem[{{Townsend} {et~al}\mbox{.}(2010){Townsend}, {Oksala}, {Cohen},
  {Owocki}, \& {ud-Doula}}]{town2010}
{Townsend} R.~H.~D., {Oksala} M.~E., {Cohen} D.~H., {Owocki} S.~P., {ud-Doula}
  A., 2010, \apjl, 714, L318

\bibitem[{{Trigilio} {et~al}\mbox{.}(2008){Trigilio}, {Leto}, {Umana}, {Buemi},
  \& {Leone}}]{2008MNRAS.384.1437T}
{Trigilio} C., {Leto} P., {Umana} G., {Buemi} C.~S., {Leone} F., 2008, \mnras,
  384, 1437

\bibitem[{{ud-Doula} {et~al}\mbox{.}(2014){ud-Doula}, {Owocki}, {Townsend},
  {Petit}, \& {Cohen}}]{ud2014}
{ud-Doula} A., {Owocki} S., {Townsend} R., {Petit} V., {Cohen} D., 2014,
  \mnras, 441, 3600

\bibitem[{{ud-Doula} \& {Owocki}(2002)}]{ud2002}
{ud-Doula} A., {Owocki} S.~P., 2002, ApJ, 576, 413

\bibitem[{{ud-Doula}, {Owocki} \& {Townsend}(2008){ud-Doula}, {Owocki}, \&
  {Townsend}}]{ud2008}
{ud-Doula} A., {Owocki} S.~P., {Townsend} R.~H.~D., 2008, MNRAS, 385, 97

\bibitem[{{ud-Doula}, {Owocki} \& {Townsend}(2009){ud-Doula}, {Owocki}, \&
  {Townsend}}]{ud2009}
{ud-Doula} A., {Owocki} S.~P., {Townsend} R.~H.~D., 2009, MNRAS, 392, 1022

\bibitem[{{ud-Doula} {et~al}\mbox{.}(2013){ud-Doula}, {Sundqvist}, {Owocki},
  {Petit}, \& {Townsend}}]{ud2013}
{ud-Doula} A., {Sundqvist} J.~O., {Owocki} S.~P., {Petit} V., {Townsend}
  R.~H.~D., 2013, \mnras, 428, 2723

\bibitem[{{Vink}, {de Koter} \& {Lamers}(2001){Vink}, {de Koter}, \&
  {Lamers}}]{vink2001}
{Vink} J.~S., {de Koter} A., {Lamers} H.~J.~G.~L.~M., 2001, \aap, 369, 574

\bibitem[{{Wade} {et~al}\mbox{.}(2015){Wade}, {Barb{\'a}}, {Grunhut},
  {Martins}, {Petit}, {Sundqvist}, {Townsend}, {Walborn}, {Alecian}, {Alfaro},
  {Ma{\'{\i}}z Apell{\'a}niz}, {Arias}, {Gamen}, {Morrell}, {Naz{\'e}}, {Sota},
  {ud-Doula}, \& {MiMeS Collaboration}}]{2015MNRAS.447.2551W}
{Wade} G.~A. {et~al.}, 2015, \mnras, 447, 2551

\bibitem[{{Wade} {et~al}\mbox{.}(2012{\natexlab{a}}){Wade}, {Grunhut},
  {Gr{\"a}fener}, {Howarth}, {Martins}, {Petit}, {Vink}, {Bagnulo}, {Folsom},
  {Naz{\'e}}, {Walborn}, {Townsend}, \& {Evans}}]{wade2012a}
{Wade} G.~A. {et~al.}, 2012{\natexlab{a}}, \mnras, 419, 2459

\bibitem[{{Wade} {et~al}\mbox{.}(2011){Wade}, {Howarth}, {Townsend}, {Grunhut},
  {Shultz}, {Bouret}, {Fullerton}, {Marcolino}, {Martins}, {Naz{\'e}}, {Ud
  Doula}, {Walborn}, \& {Donati}}]{wade2011}
{Wade} G.~A. {et~al.}, 2011, MNRAS, 416, 3160

\bibitem[{{Wade} {et~al}\mbox{.}(2012{\natexlab{b}}){Wade}, {Ma{\'{\i}}z
  Apell{\'a}niz}, {Martins}, {Petit}, {Grunhut}, {Walborn}, {Barb{\'a}},
  {Gagn{\'e}}, {Garc{\'{\i}}a-Melendo}, {Jose}, {Moffat}, {Naz{\'e}}, {Neiner},
  {Pellerin}, {Penad{\'e}s Ordaz}, {Shultz}, {Sim{\'o}n-D{\'{\i}}az}, \&
  {Sota}}]{wade2012b}
{Wade} G.~A. {et~al.}, 2012{\natexlab{b}}, \mnras, 425, 1278

\bibitem[{{Wade} \& {MiMeS Collaboration}(2015)}]{2015ASPC..494...30W}
{Wade} G.~A., {MiMeS Collaboration}, 2015, in Astronomical Society of the
  Pacific Conference Series, Vol. 494, Physics and Evolution of Magnetic and
  Related Stars, {Balega} Y.~Y., {Romanyuk} I.~I., {Kudryavtsev} D.~O., eds.,
  p.~30

\bibitem[{{Wade} {et~al}\mbox{.}(2016){Wade}, {Neiner}, {Alecian}, {Grunhut},
  {Petit}, {Batz}, {Bohlender}, {Cohen}, {Henrichs}, {Kochukhov}, {Landstreet},
  {Manset}, {Martins}, {Mathis}, {Oksala}, {Owocki}, {Rivinius}, {Shultz},
  {Sundqvist}, {Townsend}, {ud-Doula}, {Bouret}, {Braithwaite}, {Briquet},
  {Carciofi}, {David-Uraz}, {Folsom}, {Fullerton}, {Leroy}, {Marcolino},
  {Moffat}, {Naz{\'e}}, {Louis}, {Auri{\`e}re}, {Bagnulo}, {Bailey},
  {Barb{\'a}}, {Blaz{\`e}re}, {B{\"o}hm}, {Catala}, {Donati}, {Ferrario},
  {Harrington}, {Howarth}, {Ignace}, {Kaper}, {L{\"u}ftinger}, {Prinja},
  {Vink}, {Weiss}, \& {Yakunin}}]{2016MNRAS.456....2W}
{Wade} G.~A. {et~al.}, 2016, \mnras, 456, 2

\bibitem[{{Weber} \& {Davis}(1967)}]{wd1967}
{Weber} E.~J., {Davis}, Jr. L., 1967, \apj, 148, 217

\end{thebibliography}

\end{document}